\documentclass[aps,prd,showpacs,preprintnumbers,amsmath,amssymb,%
               superscriptaddress,twocolumn,notitlepage,nofootinbib]{revtex4-1}

\usepackage{graphicx} 
\usepackage{dcolumn}
\usepackage{color}
\usepackage[colorlinks=true]{hyperref}

\newcommand{\bx}{\boldsymbol{x}}
\newcommand{\bk}{\boldsymbol{k}}

\newcommand{\calO}{\mathcal{O}}

\newcommand{\be}{\begin{equation}}
\newcommand{\ee}{\end{equation}}
\newcommand{\ba}{\begin{eqnarray}}
\newcommand{\ea}{\end{eqnarray}}

\def\roughly#1{\mathrel{\raise.3ex\hbox{$#1$\kern-.75em%
\lower1ex\hbox{$\sim$}}}}

\def\slashchar#1{\setbox0=\hbox{$#1$}  
   \dimen0=\wd0     
   \setbox1=\hbox{/} \dimen1=\wd1  
   \ifdim\dimen0>\dimen1   
      \rlap{\hbox to \dimen0{\hfil/\hfil}} 
      #1     
   \else     
      \rlap{\hbox to \dimen1{\hfil$#1$\hfil}} 
      /      
   \fi}

\makeatletter
\def\overbracket#1{\mathop{\vbox{\ialign{##\crcr\noalign{\kern3\p@}
\downbracketfill\crcr\noalign{\kern3\p@\nointerlineskip}
$\hfil\displaystyle{#1}\hfil$\crcr}}}\limits}
\def\underbracket#1{\mathop{\vtop{\ialign{##\crcr
$\hfil\displaystyle{#1}\hfil$\crcr\noalign{\kern3\p@\nointerlineskip}
\upbracketfill\crcr\noalign{\kern3\p@}}}}\limits}
\def\upbracketfill{$\m@th\makesm@sh{\llap{\vrule\@height3\p@\@width.7\p@}}%
\leaders\vrule\@height.7\p@\hfill
\makesm@sh{\rlap{\vrule\@height3\p@\@width.7\p@}}$}
\def\downbracketfill{$\m@th
\makesm@sh{\llap{\vrule\@height.7\p@\@depth2.3\p@\@width.7\p@}}%
\leaders\vrule\@height.7\p@\hfill
\makesm@sh{\rlap{\vrule\@height.7\p@\@depth2.3\p@\@width.7\p@}}$}
\makeatother

\begin{document}

\title{Unruh effect and condensate in and out of an accelerated vacuum}

\author{Sanjin Beni\' c}
\affiliation{Physics Department, Faculty of Science, 
             University of Zagreb, Zagreb 10000, Croatia}
\affiliation{Department of Physics, The University of Tokyo,
             7-3-1 Hongo, Bunkyo-ku, Tokyo 113-0033, Japan}

\author{Kenji Fukushima}
\affiliation{Department of Physics, The University of Tokyo,
             7-3-1 Hongo, Bunkyo-ku, Tokyo 113-0033, Japan}
\begin{abstract}
 We revisit the Unruh effect to investigate how finite acceleration
 would affect a scalar condensate.  We discuss a negative thermal-like
 correction associated with acceleration.  From the correspondence
 between thermo-field dynamics and acceleration effects we give an
 explanation for this negative sign.  Using this result and solving the
 gap equation we show that the condensate should increase with larger
 acceleration.
\end{abstract}


\maketitle

\section{Introduction}

Thermal nature inheres in quantum field theory in spacetime with an
event horizon and it is characterized by the widely known
Hawking-Unruh temperature~\cite{Hawking:1974sw,Unruh:1976db};
$T_H = \kappa/(2\pi)$ for the black-hole case, where $\kappa = 1/(4M)$
is the surface gravity at the horizon and $M$ is the black-hole mass.
For $M\sim M_{\odot}$ (solar mass), this temperature is of the order of
$10^{-8}\,$K and it is difficult to detect any direct signal for the
Hawking radiation from astrophysical observations.  Nevertheless, it
is still a fascinating idea to seek for an analogous and more
controllable system having an event horizon.  In a laboratory setup,
the role of surface gravity may be replaced by acceleration leading to
the Unruh effect
\cite{Fulling:1972md,Unruh:1976db,Unruh:1983ms,Takagi:1986kn} (see
also Ref.~\cite{Crispino:2007eb} for a recent review).  Several
interesting ideas have been put forward to test the Hawking-Unruh
effect in a laboratory, especially concerning the condensed matter
analogue~\cite{Garay:2000jj}, strong field
systems~\cite{Greiner:1985ce,Tajima:1900zz},
lasers~\cite{Chen:1998kp,Schutzhold:2006gj,Schutzhold:2010rq,%
  Iso:2010yq,Labun:2010wf,Iso:2013sm,Steinhauer:2014dra}, and heavy
ion collisions~\cite{Castorina:2007eb,Biro:2011ne}.

The basic premise of the Unruh effect is that an accelerated observer
sees the Minkowski vacuum as a thermal (Unruh) bath.  Importantly, the
Minkowski vacuum is not necessarily empty but sometimes endowed with
condensates.  In the ground state of Quantum Chromodynamics (QCD),
that is commonly called the QCD vacuum, for instance, the chiral
condensate makes fermionic (quark) excitations gapped and the gluon
condensate arises from the trace anomaly.  In the electroweak sector
the vacuum accommodates the Higgs condensate and the Higgs phenomena
are ubiquitous in condensed matter experiments.

The basic motivation of this work is to understand if a condensate can
be modified by a finite acceleration in general.  The possible
response of a condensate to the Unruh effect is especially intriguing
from the point of view of an analogy to thermal environments;  in
ordinary thermal field theories finite temperature tends to destroy
the condensate, while the effect of finite acceleration may be or may
not be the same.

Indeed, the original works in the 1980's on the relationship between
acceleration and condensates~\cite{Hill:1985wi,Hill:1986ec} led to a
conclusion that acceleration would not have any effect on the
condensate.  This was motivated by a general theorem in interacting
field theories by Unruh and Weiss~\cite{Unruh:1983ac} on the equality
of correlations functions quantized in Rindler and Minkowski
spacetimes.

However, a number of recent works have found strikingly different
results.  The investigation of Ref.~\cite{Ohsaku:2004rv} (see also
Ref.~\cite{Kharzeev:2005iz}) on the quark-antiquark scalar condensate
(that is, the chiral condensate) in a Nambu--Jona-Lasinio model
concluded that the condensate should decrease as a function of
increasing acceleration.  A similar conclusion was reached in
Ref.~\cite{Ebert:2006bh} for a quark-quark (i.e.\ diquark) condensate
that realizes in color-superconducting phases.  In
Refs.~\cite{Lenz:2010vn,Castorina:2012yg,Takeuchi:2015nga} a real
scalar field theory was investigated in an accelerating frame, with
the main result being that a scalar condensate would decrease at
finite acceleration.  Moreover, holographic models of
QCD~\cite{Peeters:2007ti,Paredes:2008cr,Ghoroku:2010sp} indicate that
acceleration acts to weaken the interaction between quarks and
antiquarks in hadrons leading to a deconfinement transition.  In
summary, according to these preceding works, phases with a finite
scalar condensate (and also confining effects) would eventually be
destroyed at some high acceleration, in a similar way as it occurs at
high temperature.

The interpretation of the acceleration effect on condensates is far
less clear than that of the temperature effect, a part of which should
be attributed to different physical setups.  This makes it imperative
to reconsider the thermal-like effects in accelerated systems.  Thus,
in this work we revisit a real scalar field theory in Rindler
spacetime.  We explicitly compute the Wightman two-point function in
the Rindler (accelerated) vacuum.  The most subtle part is the
treatment of the ultraviolet (UV) divergence in the Rindler and the
Minkowski vacua.  To make our assumption clear, we discuss the role
played by the observer;  the observer knows the energy dispersion
relation and defines the particle there.  Then, two-point functions
involving only field operators but not the dispersion relation are
insensitive to which of the Rindler and the Minkowski vacua is chosen
for the field quantization.  Besides, our observer would not
reorganize the vacuum structure.  This means that we should treat the
UV divergence in the same way as the finite-temperature field theory,
so that we can focus only on a finite correction induced by
acceleration.  Interestingly, we will see that this
acceleration-induced correction has a sign opposite to what is
expected as a thermal correction.  We develop an analogy to the
formalism of thermo-field dynamics (TFD) to clarify the the origin of
this opposite sign.

This opposite sign reverses the role of thermal-like effects and
brings an exotic possibility that a larger condensate could be favored
with increasing acceleration.  This is the case when the condensate is
observed in the co-accelerated frame.  Utilizing the one-loop
mean-field approximation, we will solve the equation of motion and
calculate the condensate numerically and analytically.  We employ a
boundary condition that ensures that we can smoothly reach a
Minkowski-vacuum limit when the acceleration is turned off.  Our
solution exhibits divergent behavior of the condensate as a function
of the acceleration.  We might well call such a property of accelerated
matter ``acceleration catalysis'' in analogy to the magnetic
catalysis~\cite{Klimenko:1992ch}.

The nature of condensates on non-trivial spacetime
manifolds~\cite{Inagaki:1997kz,Dusling:2012ig,Yamamoto:2014vda} and in
non-inertial frames, i.e.\ accelerating or rotating
frames~\cite{Yamamoto:2013zwa}, is an interesting topic in general,
and our results should be valuable in that perspective.  Especially
interesting is the case with the Schwarzschild
metric~\cite{Flachi:2011sx} which takes the form of Rindler spacetime
near the horizon.  Hence, based on our finding we will give a brief
remark about a possible implication to the condensation phenomena in
the vicinity of the black hole.

In order to appreciate the difference from finite temperature physics,
we carefully layout the Rindler spacetime formalism and the Bogolyubov
transformation in a scalar field theory in Sec.~\ref{sec:rindsc}.
Readers familiar with this description may skip this part and proceed
directly to Sec.~\ref{sec:wigh} where we discuss the basic expressions
about field correlations and number operators with acceleration.  Also
we discuss the correspondence between the accelerated vacuum and the
thermal vacuum.  Such a careful comparison provides us with a key to
the phenomena of acceleration catalysis which is addressed in
Sec.~\ref{sec:ssb}.  We make our conclusions in Sec.~\ref{sec:concl}.
We give an explicit check of the equality between the Wightman
two-point functions in the Rindler and the Minkowski vacua in
Appendix.

\section{Inequivalent vacua and observers}
\label{sec:observers}

We would stress the importance to sort out the definitions of the
\textit{vacua} (or \textit{states}) and the observers (or
\textit{operators}) first.  Let us start our discussions with an
analogous and more intuitive example of particle production under an
external electric field, i.e.\ the phenomenon called the Schwinger
mechanism~\cite{Dunne:2004nc}.  The problem with an electric field is
essentially dynamical in a sense that the background gauge fields
should be time dependent.  It is convenient to introduce quantities
and operators in the infinitely past (and future) state that are
referred to with a subscript ``in'' (and ``out'' respectively).

It is a well-known result that the in-vacuum is not really a vacuum if
seen by an observer sitting in the out-vacuum, which is explicitly
expressed in a form of the in-state expectation value of the out-state
number operator.  For example, if the electric field is applied for a
time $\sim t$ along the positive $z$-direction, the production of
charged bosonic particles results in a distribution as
follows~\cite{Fukushima:2009er}:
\begin{equation}
 \begin{split}
  & \langle\text{in}| \hat{a}_{\text{out}}^\dag(\bk)
  \hat{a}_{\text{out}}(\bk)|\text{in}\rangle \\
  &\qquad\quad \sim
  \exp\biggl[-\frac{\pi(\bk_\perp^2+m^2)t}{4}\Bigl(
  \frac{1}{k^z-eEt} + \frac{1}{k^z}\Bigr)\biggr]
 \end{split}
\label{eq:Schwinger}
\end{equation}
for $0\leq k^z\leq eEt$.  This non-zero result appears from the
Bogolyubov coefficients between $\hat{a}_{\text{in}}$,
$\hat{a}_{\text{in}}^\dag$ and $\hat{a}_{\text{out}}$,
$\hat{a}_{\text{out}}^\dag$.

In Eq.~\eqref{eq:Schwinger} the in-vacuum $|\text{in}\rangle$ is
probed by an out-operator.  In other words, the observer
\textit{defines} the operator we should put in the expectation value.
To understand this machinery more, it would be instructive to recall
how the number operator can be written in terms of field operators.
As derived in Ref.~\cite{Fukushima:2014sia}, we can show:
\begin{equation}
 \begin{split}
  & \hat{a}_{\text{out}}^\dag(\bk)\hat{a}_{\text{out}}(\bk)
    = \frac{1}{2\varepsilon_{\text{out}}(\bk)}\lim_{t_1=t_2\to\infty} \\
  & \times [\partial_{t_1}+i\varepsilon_{\text{out}}(\bk)]
           [\partial_{t_2}+i\varepsilon_{\text{out}}(\bk)]
           \hat{\phi}^\dag(t_1,\bk)\hat{\phi}(t_2,\bk)\;,
 \end{split}
\end{equation}
where $\varepsilon_{\text{out}}(\bk)$ is the energy dispersion
relation in the out-state which generally depends on when and where
the particle is observed.  Importantly, as we will explicitly confirm
later, the field operators, $\hat{\phi}$ and $\hat{\phi}^\dag$, are
not sensitive to the detection procedures.  A more familiar and
general example of the relevance of the out-observer through the
energy dispersion relation can be also found in the famous LSZ
(Lehmann-Symanzik-Zimmermann) reduction formula.

Now, we shall turn to the consideration about the Unruh effect.  It is
crucial to make the observer's role clear in order to clarify the
physical interpretation of the Unruh effect.  We will see later that
the non-accelerated vacuum expectation value of the number operator in
the accelerated vacuum, $\langle M|\hat{a}_R^\dag \hat{a}_R|M\rangle$,
has a thermal spectrum whose temperature is characterized by the
acceleration.  Then, one might be tempted to consider that this
non-accelerated vacuum could be a thermal bath giving rise to
thermal-like corrections.  Such an argument would cause confusion if
applied too na\"{i}vely, and the fact is that there is no such
thermal-like correction as long as an operator is written in terms of
$\hat{\phi}$ and $\hat{\phi}^\dag$ only and not with the energy
dispersion relation inherent in the observer.

Nevertheless, even for operators not involving the energy dispersion
relation, it is still a non-trivial question how the operator
expectation value may change with different vacua; the non-accelerated
vacuum $|M\rangle$ and the accelerated one $|R\rangle$.  This is a
question that we elucidate in the present work.  In particular, we are
interested in a scalar condensate affected by the acceleration.  In
summary, we will take a close look at the question:
\begin{equation}
 \langle M|\hat{\phi}|M\rangle \overset{?}{=}
 \langle R|\hat{\phi}|R\rangle
\end{equation}
and think about underlying physical interpretations in analogy to
thermal field theory in what follows below.

\section{Unruh Effect in a Scalar Field Theory}
\label{sec:rindsc}

This is an overview section and we summarize our notations and choice
of the coordinates, i.e.\ those in Rindler spacetime.  These
preliminary setups are important for the later analysis on the
spontaneous symmetry breaking in Sec.~\ref{sec:ssb}.

\subsection{Scalar field in Rindler spacetime}

The Minkowski metric is given as $ds^2 = dt^2 - d\bx_\perp^2 - dz^2$
in our convention where $\bx_\perp = (x,y)$.  We perform a change of
coordinate variables from $t$ and $z$ to $\rho$ and $\eta$, which
defines the Rindler coordinates as follows:
\be
  z = \rho \cosh \eta ~, \qquad t = \rho \sinh \eta
\label{eq:rindtr}
\ee
with the metric in a form of
\be
  ds^2 = \rho^2 d\eta^2 - d\rho^2 - d\bx_\perp^2\;.
\label{eq:rindm}
\ee
These new coordinates, $\rho$ and $\eta$, cover only a part of the
Minkowski spacetime as long as $\rho$ is non-negative and $\eta$ is
real.  Because the region of $z>|t|$ is spanned then, we call $\rho$
and $\eta$ the \textit{right-wedge} Rindler coordinates.  We can also
introduce another coordinates, $\bar{\rho}$ and $\bar{\eta}$, to
define the \textit{left-wedge} Rindler coordinates in a similar
fashion.  In this work we will focus on the right-wedge Rindler
coordinates only;  we can setup an accelerated particle trajectory
within this $z>0$ region without loss of generality.

\begin{figure}
 \includegraphics[width=0.8\columnwidth]{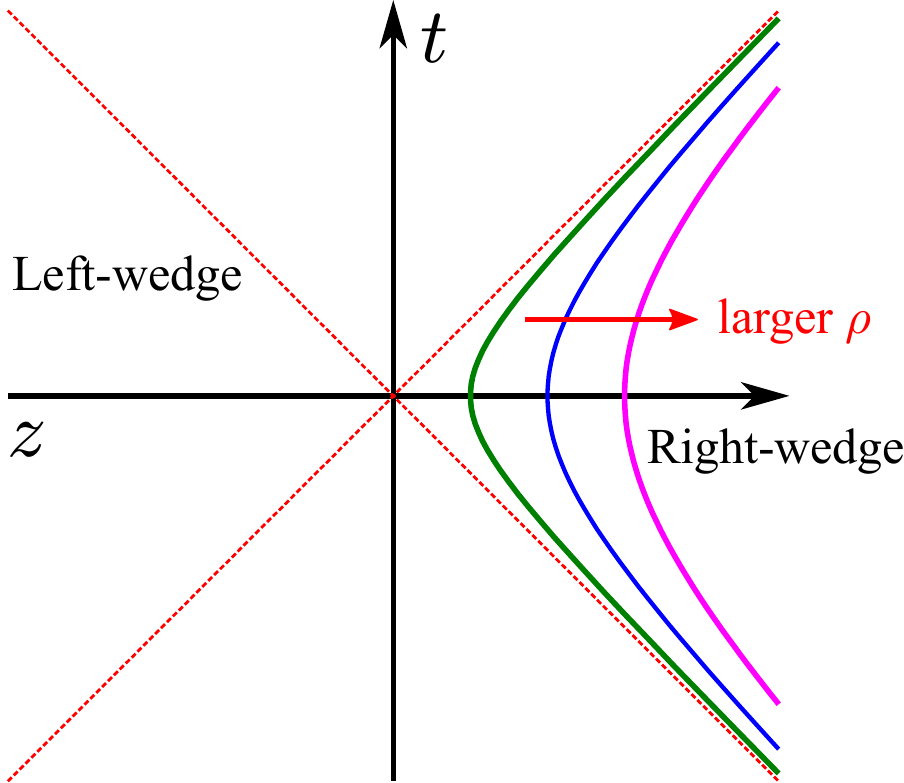}
 \caption{Schematic picture of the Rindler coordinates.  In the
   right-wedge Rindler coordinates several curves with different values
   of $\rho$ are shown;  a larger $\rho$ makes the constant-$\rho$
   trajectory straightened corresponding to a less acceleration.}
 \label{fig:spacetime}
\end{figure}

Using the notion of the proper time $\tau$ and the velocity
four-vector $u^\mu=dx^\mu/d\tau$, we can define the acceleration
four-vector as $a^\mu = d u^\mu/d\tau$.  Then, the proper acceleration
$\alpha$ is given by
\be
 \alpha^2 = -a_\mu a^\mu~.
\ee
A trajectory of a point particle with a proper acceleration $\alpha$
in terms of the Minkowski coordinates can be parametrized as
\be
 z(\tau) = \frac{1}{\alpha}\cosh(\alpha\tau)\;, \qquad 
 t(\tau) = \frac{1}{\alpha}\sinh(\alpha\tau)\;.
\ee
Therefore, in terms of the right-wedge Rindler coordinates, this
trajectory corresponds to $\eta=\alpha\tau$ with a fixed value of
$\rho=1/\alpha$.  Thus, we should stress here that $\rho$ and $\eta$
have dual roles as coordinates and parameters characterizing an
accelerated trajectory.  As sketched in Fig.~\ref{fig:spacetime}, the
constant-$\rho$ trajectories move away from the light-cone, and their
shape straightens, as $\rho$ increases.  This clearly means that a
larger $\rho$ represents a smaller acceleration, which is consistent
with the identification of $\rho=1/\alpha$.

The action for a real scalar field theory in a general coordinate
system (apart from the curvature) \cite{Parker:2009} reads:
\be
 S = \int d^4 x \sqrt{-g} \biggl[ \frac{1}{2}g^{\mu\nu}\partial_\mu \phi
  \partial_\nu \phi -V(\phi) \biggr]\;,
\ee
where $V(\phi)$ is a potential term.

Let us now setup two distinct observers; a Minkowski (non-accelerated)
observer and a Rindler (accelerated) observer.  The Minkowski and the
Rindler observers quantize the fields based on the energy dispersion
relation in Minkowski and Rindler space, respectively.  We will
indicate observables quantized in this way by denoting $\hat{\calO}_M$
and $\hat{\calO}_R$ for the Minkowski and the Rindler observers.  As
we already mentioned above, this distinction is irrelevant for
observables in terms of $\hat{\phi}$ and $\hat{\phi}^\dag$ only.

We briefly see how the Unruh effect is derived.  With the Minkowski
coordinates the quantum field is expanded in terms of a complete set
of basis functions (denoted by $f$'s) with the creation and
annihilation operators as
\be
 \hat{\phi} = \int d^3 k \Bigl[\hat{a}_M(\bk) f(\bk,x)
  + \hat{a}_M^\dag(\bk) f^\ast(\bk,x)\Bigr] \;.
\label{eq:phim}
\ee
The choice of the complete set is arbitrary and it would be the most
convenient one to take the plane waves as the basis functions as
\be
 f(\bk,x) = \frac{1}{(2\pi)^{3/2}(2k_0)^{1/2}}\,
  e^{i\bk\cdot \bx-ik_0 t}\;.
\ee
We thus define the Minkowski vacuum $|M\rangle$ as a solution of
$\hat{a}_M(\bk)|M\rangle=0$.

With the Rindler metric~\eqref{eq:rindm} we should replace $f(\bk,x)$
with a counterpart of the plane wave in terms of the right-wedge
Rindler coordinates (see, e.g.\ 
Refs.~\cite{'tHooft:1996tq,Crispino:2007eb,Lenz:2010vn} for technical
details) given by
\be
 \begin{split}
 & f_R(\bk_\perp,\omega,x) \\
 &\qquad  = \frac{\sqrt{1-e^{-2\pi\omega}}}{[2(2\pi)^4]^{1/2}}
 K\Bigl(\omega,\frac{\kappa\rho}{2},
 \frac{\kappa\rho}{2}\Bigr)\,e^{i\bk_\perp\cdot \bx_\perp-i\omega\eta}\;,
 \end{split}
\label{eq:reigen}
\ee
where we introduced $\kappa\equiv\sqrt{\bk_\perp^2+m^2}$.  We note
that $\omega$ is a dimensionless conjugate of $\eta$.  From the
trajectory, $\eta=\alpha\tau$, we should understand that
$\alpha\omega$ corresponds to a Rindler energy.  The $\rho$ dependence
appears through a special function defined as
\be
 K(\omega,\alpha,\beta) \equiv \int_0^\infty \frac{ds}{s} s^{i\omega}\,
 e^{-is\alpha + i\beta/s}\;.
\label{eq:bessi}
\ee
We should consider this function in the physical region of $\omega>0$
where we can check the following:
\begin{equation}
 K(\omega,\alpha,\alpha) = 2e^{\pi\omega/2}K_{i\omega}(2\alpha)\;.
\label{eq:bessrel}
\end{equation}
Here, $K_{i\omega}(x)$ represents the modified Bessel function of
imaginary order.  We can then expand the quantum field for the Rindler
observer as
\be
\begin{split}
 \hat{\phi} &= \int_0^{\infty}d\omega\int d^2k_\perp \Bigl[
  \hat{a}_R(\bk_\perp,\omega)\,f_R(\bk_\perp,\omega,x) \\
  &\qquad\qquad\qquad\qquad\quad
  +\hat{a}_R^\dag(\bk_\perp,\omega)\,f_R^\ast(\bk_\perp,\omega,x)\Big]\;.
\label{eq:phir2}
\end{split}
\ee
In the same way as for Minkowski spacetime, we can define the vacuum
for the Rindler observer by solving
$\hat{a}_R(\bk_\perp,\omega)|R\rangle = 0$.  It is also possible to
perform an expansion of the field in the left-wedge to construct
another vacuum from $\hat{a}_L(\bk_\perp,\omega)|L\rangle = 0$ if
necessary.

\subsection{Bogolyubov transformation and the Unruh temperature}

We establish the relation between $\hat{a}_M(\bk)$ and
$\hat{a}_{R/L}(\bk_\perp,\omega)$, which is formulated conveniently in
terms of the Bogolyubov coefficients.  Even though the
transformation~\eqref{eq:rindtr} is just a change of variables, the
existence of a causal horizon at $t=\pm z$ for the accelerated
observer introduces a non-trivial structure through the Bogolyubov
transformation and this is at the heart of the Unruh effect.

Following Ref.~\cite{'tHooft:1996tq} we first define a light-cone
annihilation operator as
$\sqrt{k^+}\hat{a}_1(\bk_\perp,k^+) = \sqrt{k_0} \hat{a}_M(\bk)$ with
$k^\pm \equiv (k^z\pm k^0)/\sqrt{2}$.  Then, we find another operator
by a variable change from $k^+$ to $\omega$, i.e.
\be
 \hat{a}_2(\bk_\perp,\omega) = \int_0^\infty\! \frac{dk^+}{(2\pi k^+)^{1/2}}\,
 \hat{a}_1(\bk_\perp,k^+)\,e^{i\omega\log[(k^+\sqrt{2})/\kappa]}\;.
\label{eq:a21}
\ee
For $\omega>0$ these new operators $\hat{a}_2(\bk_\perp,\pm\omega)$
are related to the Rindler operators $\hat{a}_{R/L}(\bk_\perp,\omega)$
through a linear transformation expressed as
\be
\begin{pmatrix}
 \hat{a}_R(\bk_\perp,\omega)\\
 \hat{a}_L(\bk_\perp,\omega)\\
 \hat{a}_R^\dag(-\bk_\perp,\omega)\\
 \hat{a}_L^\dag(-\bk_\perp,\omega)
\end{pmatrix}
=
\begin{pmatrix}
\alpha_\omega & 0 &  0 & \beta_\omega\\
0 & \alpha_\omega & \beta_\omega &  0\\
0 & \beta_\omega & \alpha_\omega & 0\\
\beta_\omega & 0 & 0 & \alpha_\omega
\end{pmatrix}
\begin{pmatrix}
 \hat{a}_2(\bk_\perp,\omega)\\
 \hat{a}_2(\bk_\perp,-\omega)\\
 \hat{a}_2^\dag(-\bk_\perp,\omega)\\
 \hat{a}_2^\dag(-\bk_\perp,-\omega)
\end{pmatrix}
\label{eq:bog}
\ee
with the Bogolyubov coefficients given by
\be
 \alpha_\omega = \frac{1}{\sqrt{1-e^{-2\pi\omega}}} \;,
 \qquad
 \beta_\omega = \frac{e^{-\pi\omega}}{\sqrt{1-e^{-2\pi\omega}}} \;.
\label{eq:beta}
\ee
Using the above Bogolyubov transformation and the operator
definition~\eqref{eq:a21}, we can readily see that the Minkowski
vacuum expectation value of the Rindler number operator is non-zero,
namely,
\begin{align}
 &\langle M|\hat{a}_R^\dag(\bk_\perp,\omega)
  \hat{a}_R(\bk_\perp',\omega')|M\rangle \notag\\
 &= \beta_\omega\beta_{\omega'}\delta^{(2)}(\bk_\perp\!-\!\bk_\perp')
  \int_0^\infty \!\frac{dk^+}{2\pi k^+}\,
  e^{-i(\omega-\omega')\log[(k^+\sqrt{2})/\kappa]} \notag\\
 &= \frac{\delta(\omega-\omega')
  \delta^{(2)}(\bk_\perp-\bk_\perp')}{e^{\omega/T_U}-1} \;,
\label{eq:therm1}
\end{align}
which demonstrates the Unruh effect~\cite{Unruh:1976db}, where
$T_U = 1/(2\pi)$ is the dimensionless Unruh temperature.  (In the
physical unit, $\alpha\omega$ is a Rindler energy, and so
$\alpha/(2\pi)$ is to be identified as the temperature.)  One may be
tempted to interpret $|M\rangle$ as a thermal bath for operators
quantized on the Rindler vacuum.  Such an argument for the thermal
interpretation could be found in some literature, but it is sometimes
concluded in a rather misleading manner.  We will elucidate this point
in more details in the next section.

\section{Calculating Thermal-like Corrections}
\label{sec:wigh}

In this section we shall consider the effect of quantum fluctuations
in and out of an accelerated vacuum.  For this purpose we take an
example of two-point Wightman function that is necessary for the
evaluation of the scalar condensate.  In particular, the coincidence
limit of the two-point functions,
$\langle R|\hat{\phi}^2|R\rangle$ and
$\langle M|\hat{\phi}^2|M\rangle$, will introduce a ``temperature''
dependent mass term in the effective potential~\cite{Dolan:1973qd}.

\subsection{Insensitivity to the observer}

Although
$\langle M|\hat{a}_R^\dag \hat{a}_R|M\rangle\neq
 \langle M|\hat{a}_M^\dag \hat{a}_M|M\rangle=0$ and
$\langle R|\hat{a}_M^\dag \hat{a}_M|R\rangle\neq
 \langle R|\hat{a}_R^\dag \hat{a}_R|R\rangle=0$, the Bogolyubov
coefficients guarantee that $\hat{\phi}$ represents the same quantum
field.  In fact, for $\calO(\hat{\phi})$ not having the energy
dispersion relations, it has been addressed based on the functional
integration in the literature~\cite{Unruh:1983ac} that
$\langle M|\calO(\hat{\phi})|M\rangle$ does not depend on the choice
of the observer who quantizes $\hat{\phi}$.  Moreover,
Refs.~\cite{Hill:1985wi,Hill:1986ec} utilized the Schr\"odinger
functional formalism with an explicit point-splitting regularization
to prove that $\langle M|\calO(\hat{\phi})|M\rangle$ should be
independent of the observer.

This is all so by construction, and nevertheless, operators for
different observers (i.e.\ quantized in different vacua) sometimes
cause confusions.  Thus, it would be useful to take a glance at how
the insensitivity follows explicitly from a proper combination of
the vacuum definitions and the Bogolyubov coefficients.  The
calculation to confirm the insensitivity of both
$\langle M|\calO(\hat{\phi})|M\rangle$ and
$\langle R|\calO(\hat{\phi})|R\rangle$ for different observers is
tedious but straightforward.  This is a two-step procedure to use the
Bogolyubov relations~\eqref{eq:bog} and the transformation of plane
waves to
\be
\begin{split}
 &e^{i(k^+ z^- + k^- z^+)} \\
 &= \int_{-\infty}^{\infty}\frac{d\omega}{2\pi}\,
  e^{-i\omega\eta}e^{i\omega\log(k^+\sqrt{2}/\kappa)}
  K\Bigl(\omega,\frac{\kappa\rho}{2},\frac{\kappa\rho}{2}\Bigr)~,
\end{split}
\label{eq:usef}
\ee
and vice versa.  We give more detailed and complete calculations in
the Appendix~\ref{sec:app}.

\subsection{Regularization prescription}
\label{sec:therli}

We will explicitly evaluate the coincidence limit $x'\to x$ of
two-point Wightman functions for $m=0$.  For concrete calculations we
could use the point-splitting regularization.  For the $m=0$ case,
then, we have:
\be
\begin{split}
 &\langle M|\hat{\phi}(x)\hat{\phi}(x')|M\rangle \\
 & = -\frac{1}{4\pi}\frac{1}{(t-t'-i\epsilon)^2
  - |\bx_\perp-\bx_\perp'|^2 - |z-z'|^2}~.
\end{split}
\label{eq:mindiv}
\ee
For the Rindler vacuum also the Wightman function
$\langle R|\hat{\phi}(x)\hat{\phi}(x')|R\rangle$ has a singular term
in the coincident limit just given by Eq.~\eqref{eq:mindiv} (see also
Refs.~\cite{Hill:1985wi,Hill:1986ec}) as well as a finite deviation.
Such a finite extra term should be well-defined irrespective to the
ultraviolet regularization.  We can actually find:
\begin{align}
 &\langle R|\hat{\phi}^2|R\rangle_{\rm reg} \notag\\
 &\equiv \lim_{x'\to x}\bigl( \langle R|\hat{\phi}(x)\hat{\phi}(x')|R \rangle
 - \langle M|\hat{\phi}(x)\hat{\phi}(x')|M\rangle \bigr) \notag\\
 &= -\frac{1}{4\pi^4}
  \int_0^\infty d\omega\, e^{-\pi\omega} \int d^2 k_\perp
  K_{i\omega}^2(k_\perp\rho) \notag\\
 &= -\frac{1}{2\pi^2\rho^2}\int_0^\infty
  \frac{\omega d\omega}{e^{2\pi\omega}-1}
  = -\frac{1}{48\pi^2 \rho^2}
\label{eq:corrr}
\end{align}
for the $m=0$ case.  Recalling that the trajectory of a constant
proper acceleration $\alpha$ is defined as $\rho = 1/\alpha$ and
defining the local Unruh temperature as
$T_{\rm loc} = \alpha/(2\pi)$, we can interpret this result as a
thermal-like correction by
\be
 \langle R|\hat{\phi}^2|R\rangle_{\rm reg} = -\frac{T_{\rm loc}^2}{12}~.
\label{eq:depre1}
\ee
It should be noted that this expression~\eqref{eq:depre1} has a sign
opposite to the ordinary thermal correction if $|R\rangle$ is given an
interpretation as a thermal
bath~\cite{Candelas:1976jv,Hill:1985wi,Hill:1986ec,Stephens:1986ie,%
Lenz:2010vn}.

Before closing this subsection, we make a remark that in the canonical
quantization it is a conventional procedure to take the normal
ordering to discard zero-point oscillation energies, that is,
\begin{equation}
 :\hat{\calO}_{R/M}: \equiv \hat{\calO}_{R/M}
  - \langle R/M|\hat{\calO}_{R/M} |R/M \rangle\;, 
\end{equation}
where the second contribution represents the discarded divergent piece
in the normal ordering in terms of $\hat{a}_R$ and $\hat{a}_R^\dag$
and in terms of $\hat{a}_M$ and $\hat{a}_M^\dag$, respectively.  In
this case, even when $\calO_{R/M}$ has no explicit dependence on the
energy dispersion relation, the expectation value may change according
to the observer through $\langle R/M|\hat{\calO}_{R/M} |R/M \rangle$.
It is obvious that the subtraction \eqref{eq:corrr} coincides with
the normal ordering for the Minkowski observer.

Because we have no complete description of the zero-point oscillation
but dropping it with some working prescriptions in quantum field
theory, we should choose a reference point where we make a subtraction
of the divergent term as in Eq.~\eqref{eq:corrr}.  As emphasized in
Sec.~\ref{sec:observers} our assumption about the observer is that the
observer defines the energy dispersion relation that is needed for
switching to the particle picture.  Hence, in our prescription, the
observer \textit{does not reorganize} the vacuum, and so an offset of
the energy level should be intact.  This assumption thus prescribes us
not to include this difference between
$\langle R|\hat{\calO}_R|R\rangle$ and $\langle
M|\hat{\calO}_M|M\rangle$ in our computation.  In a sense our
treatment of the UV singularity is analogous to that in
finite-temperature field theory;  once divergences are subtracted at
$T=0$, no additional divergence appears from $T\neq0$ corrections.
More specifically, for a thermal state $|\beta\rangle$ with
temperature $T$, for a free massless scalar theory, we know:
\begin{align}
 \langle\beta| \hat{\phi}^2 |\beta\rangle_{\rm reg}
 &\equiv \lim_{x' \to x} \bigl(
   \langle\beta| \hat{\phi}(x)\hat{\phi}(x') |\beta\rangle
  -\langle 0| \hat{\phi}(x)\hat{\phi}(x') |0\rangle \bigr) \notag\\
 &= \frac{T^2}{12}\;,
\label{eq:threg}
\end{align}
which is quite suggestive as compared to Eq.~\eqref{eq:depre1}.  We
shall pursue for this analogy to finite-$T$ field theory more in the
next subsection.

\subsection{Analogue to thermo-field dynamics}

We see a clear correspondence from expectation values of the number
operator in Rindler spacetime and in thermal environments.  Indeed, a
striking similarity is found, which takes the form of
\be
 \langle \beta|\hat{a}_0^\dag(\bk) \hat{a}_0(\bk')|\beta\rangle
 = \frac{\delta^{(3)}(\bk-\bk')}{e^{\sqrt{\bk^2+m^2}/T}-1} \;,
\label{eq:bed}
\ee
where $\hat{a}_0$ and $\hat{a}_0^\dag$ are the annihilation and
creation operators in Minkowski spacetime.

The comparison with thermo-field dynamics (TFD) will provide us with
an intuitive understanding of the results~\eqref{eq:corrr} and
\eqref{eq:depre1}.  In TFD one deals with the thermal vacuum
$|\beta\rangle$ which is represented by the so-called non-tilde
$|0\rangle$ and tilde $|\tilde{0}\rangle$ vacua~\cite{Umezawa:1982nv}
and so $|\beta\rangle$ is excited relative to $|0\rangle$.

\begin{table}
\begin{center}
 \begin{tabular}{ccc} \hline
 TFD &$\;\Leftrightarrow\;$& Rindler Spacetime\\
 \hline\hline
 $|\beta\rangle$     && $|M\rangle$\\
 $|0\rangle$         && $|R\rangle$\\
 $|\tilde{0}\rangle$ && $|L\rangle$\\
 \hline
 \end{tabular}
\end{center}
\caption{Correspondence between different vacua in TFD and in Rindler
  spacetime.}
\label{tab:corresp}
\end{table}

It is important to recognize that $|M\rangle$ should be ``thermal'' in
terms of $|R\rangle$ and $|L\rangle$ (right-wedge and left-wedge
Rindler vacua that never talk to each other), so that $|M\rangle$ in Rindler spacetime is analogous to
$|\beta\rangle$ in TFD and $|R\rangle$ and $|L\rangle$ should
correspond to $|0\rangle$ and $|\tilde{0}\rangle$.  We summarize the
relation among them in Table~\ref{tab:corresp}.  Confusions sometimes
arise from misidentification of $|R\rangle$ as a thermal mixed state,
but the fact is opposite.  This point is important to understand the
meaning of the negative sign in Eq.~\eqref{eq:depre1}.


In TFD, the observer quantizes operators in a zero temperature vacuum
and can measure $\langle 0|\hat{\calO}|0\rangle$ to take it as a
``reference'' value.  Now, let us suppose that we have box
representing a piece of material heated to non-zero $T$.  This box is
in a thermal state $|\beta\rangle$ and the same observer should find:
\be
 \langle \beta|\hat{\calO}|\beta\rangle \;>\;
 \langle 0|\hat{\calO}|0\rangle
\label{eq:ineq_tfd}
\ee
for a positive definite operator $\hat{\calO}$ such as
$\hat{\calO}=\hat{\phi}^2$.  The left-hand side in the above receives a
thermal correction $\sim T^2$, which is a finite correction associated
with the temperature effect.

Now, let us imagine a similar experiment where the vacuum state in the
box is not heated but accelerated.  In the case with acceleration,
according to Table~\ref{tab:corresp}, we should anticipate:
\be
 \langle M|\hat{\calO}|M\rangle \;>\;
 \langle R|\hat{\calO}|R\rangle
\label{eq:ineq}
\ee
as a counterpart of the relation~\eqref{eq:ineq_tfd}.  We should take
the left-hand side in the above as our reference point before
acceleration, so that a finite correction associated with the
acceleration effect is naturally negative.  In other words, we can say
that the accelerated vacuum is \textit{less excited} as compared to
the non-accelerated (Minkowski) vacuum.

\section{Spontaneous Symmetry Breaking}
\label{sec:ssb}

We are considering a real scalar field theory with $Z_2$ symmetry,
which is assumed to be spontaneously broken in the Minkowski vacuum
through the potential of the following form:
\be
 V(\phi) = -\frac{\mu^2}{2}\phi^2 + \frac{\lambda}{4}\phi^4\;.
\label{eq:pot}
\ee
On the tree level, the state with a minimal energy favors a finite
condensate given by
\be
 \langle M|\hat{\phi}|M\rangle = \sqrt{\frac{\mu^2}{\lambda}}\;.
\label{eq:homog}
\ee
The question we are addressing in this section is the following;  let
us consider a box of, say, a superconducting material with a non-zero
homogeneous condensate such as in Eq.~\eqref{eq:homog}.  Then, we
accelerate this box and adiabatically change $|M\rangle$ to
$|R\rangle$ to investigate whether the condensate may increase or
decrease with acceleration.

\subsection{One-loop equation of motion}

We here introduce a notation;
$\bar{\phi} \equiv \langle R|\hat{\phi}|R\rangle$ and we can determine
$\bar{\phi}$ by the condition to extremize the effective action.  The
one-loop calculation in the mean-field approximation leads to the
follow equation of motion:
\be
 \bigl( \square - 3\lambda \langle\hat{\phi}^2\rangle_{\text{reg}}
  \bigr) \bar{\phi} - V'(\bar{\phi}) = 0~.
\label{eq:cback}
\ee
This is a non-linear equation for $\bar{\phi}$ involving quantum
fluctuations encoded in
$\langle \hat{\phi}^2\rangle_{\text{reg}}=\langle R|\hat{\phi}^2|R\rangle
-\langle M|\hat{\phi}^2|M\rangle$.  We should note that our reference
point is the Minkowski vacuum and so
$\langle\hat{\phi}^2\rangle_{\text{reg}}<0$.

Our goal is to find $\bar{\phi}$ as a function of $\rho$, where $\rho$
is the Rindler coordinate as introduced before.  In contrast to finite
temperature physics where the effective \textit{potential} is
sufficient to fix a condensate, in the acceleration case the
Hamiltonian depends on acceleration through the coordinate $\rho$ and
so it is indispensable to keep the derivatives to find $\bar{\phi}$ in
the accelerated vacuum.

We can adopt the quantum fluctuation
$\langle\hat{\phi}^2\rangle_{\text{reg}}$ from Sec.~\ref{sec:rindsc}
and approximately use Eq.~\eqref{eq:corrr}.  The present theory is not
a massless one, but this massless approximation would simplify the
analysis significantly not losing qualitative features.  Assuming a
non-trivial $\rho$ dependence in the condensate, the problem boils
down to solving the following equation:
\be
 \frac{d^2\bar{\phi}}{d\rho^2}+\frac{1}{\rho}\frac{d\bar{\phi}}{d\rho}
  -\nu^2 \frac{\bar{\phi}}{\rho^2} = V'(\bar{\phi})~, \quad
 \nu^2 = -\frac{\lambda}{16\pi^2}\;.
\label{eq:diff1}
\ee
Solving Eq.~\eqref{eq:diff1} allows for a particle-like
interpretation;  $\bar{\phi}$ is to be interpreted as ``position'' of
a particle and $\rho$ as ``time''.  Then, this identification enables
us to rewrite Eq.~\eqref{eq:diff1} in the ``energy'' form:
\be
 \frac{d}{d\rho}\biggl[ \frac{1}{2}
  \Bigl(\frac{d\bar{\phi}}{d\rho}\Bigr)^2 - V(\bar{\phi})\biggr]
 = -\frac{1}{\rho}\Bigl(\frac{d\bar{\phi}}{d\rho}\Bigr)^2
  + \nu^2\frac{d\bar{\phi}}{d\rho}\frac{\bar{\phi}}{\rho^2}~,
\label{eq:part}
\ee
which gives us an interpretation that a particle is moving in a
potential $-V(\bar{\phi})$.  It is crucial to point out that both
terms in the right-hand side are negative (if $d\bar{\phi}/d\rho>0$),
leading to an energy loss as a function of time.

This type of analysis is typical for the calculation of false vacuum
decay~\cite{Coleman:aos} when a potential energy has several
inequivalent minima.  One is then interested in finding an
``instanton'' solution that represents a trajectory from one to the
other extrema of the potential $-V(\bar{\phi})$.  The solution of our
current interest should satisfy a boundary condition:
\be
 \bar{\phi}(\rho\to\infty) = \sqrt{\frac{\mu^2}{\lambda}}\;,
\label{eq:bc1}
\ee
so that the condensate is reduced to its vacuum value when the proper
acceleration $\alpha = 1/\rho$ is turned off.  In the opposite limit,
one might have been tempted to impose $\bar{\phi}(\rho\to 0)=0$,
leading to a picture of symmetry restoration induced by high
acceleration.  It is obvious, however, that such a boundary condition
is incompatible, which is understood from Eq.~\eqref{eq:part} that
cannot increase the energy of the particle.  One might then think that
a boundary condition such as $(d\bar{\phi}/d\rho)_{\rho=0} > 0$ could
work to lead to a consistent solution.  However, we can easily check
by linearizing Eq.~\eqref{eq:diff1} around $\bar{\phi} = 0$ to find
that the resulting trajectories are
\be
 \bar{\phi}(\rho) = C_1\rho J_{\sqrt{1+\nu^2}}(\mu\rho)+
  C_2\rho N_{\sqrt{1+\nu^2}}(\mu\rho)~,
\ee
which have zero gradient at $\rho=0$.  Therefore, using the boundary
condition like $(d\bar{\phi}/d\rho)_{\rho=0} > 0$ would end up with an
inconsistency.

\subsection{Acceleration catalysis}
\label{ssec:acat}

\begin{figure}
\begin{center}
\includegraphics[clip,width=0.9\columnwidth]{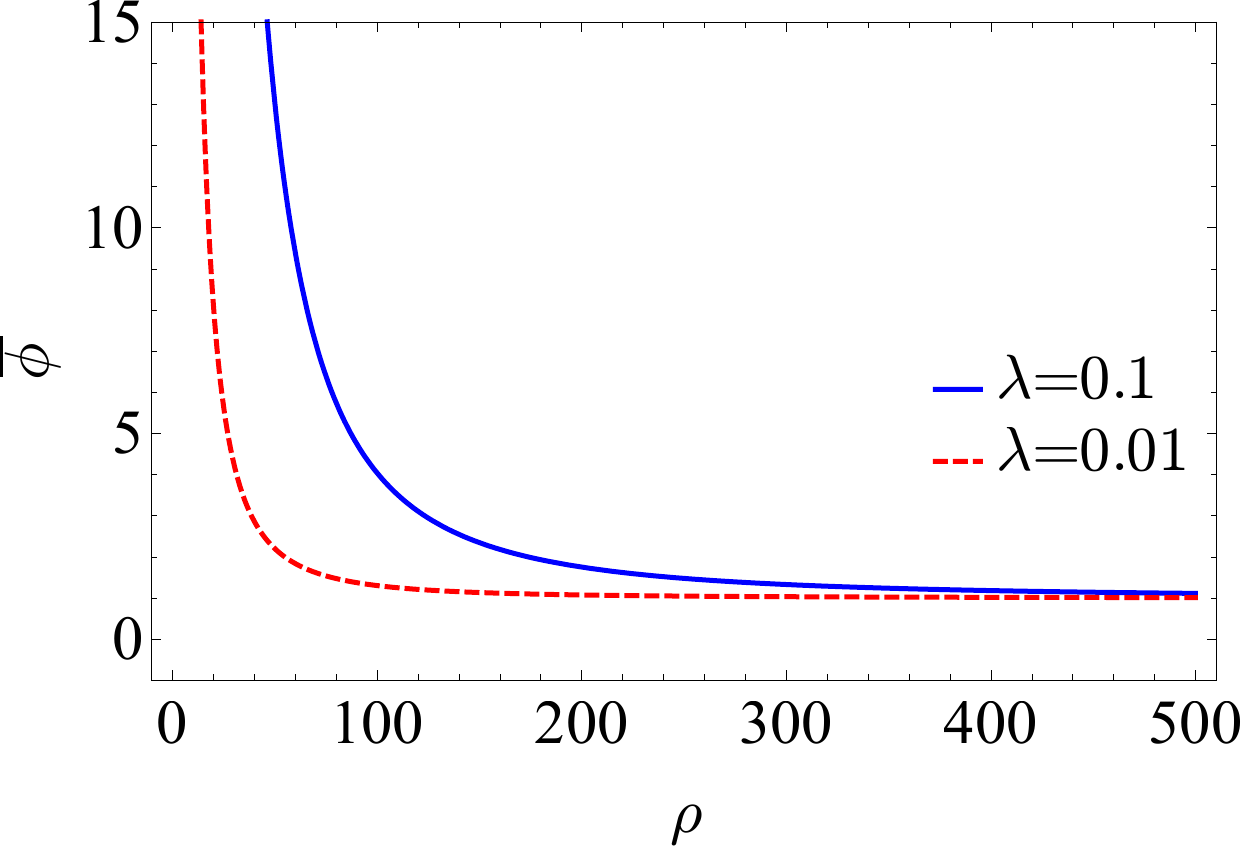}
\caption{One-loop corrected background field for $\lambda=0.1$ and
  $\lambda=0.01$ as a function of $\rho=1/\alpha$ where $\alpha$ is
  the proper acceleration.  The axes are given in dimensionless units,
  where $\bar{\phi}$ and $\rho$ are scaled by $\sqrt{\mu^2/\lambda}$.}
\label{fig:num}
\end{center}
\end{figure}

For this problem, it would be a more reasonable choice to impose one
more boundary condition at $\rho\to\infty$.  Besides
Eq.~\eqref{eq:bc1} we also require,
\be
 \biggl( \frac{d\bar{\phi}}{d\rho} \biggr)_{\rho\to\infty} = 0~,
\label{eq:bc2}
\ee
so that in the limit of zero acceleration the solution smoothly
approaches the value of the condensate in the Minkowski vacuum.

It is straightforward to see that Eq.~\eqref{eq:diff1} with $\mu=0$
accommodates a ``scaling'' type of the solution:
\be
 \bar{\phi}(\rho) = \sqrt{\frac{1+\nu^2}{\lambda}}\frac{1}{\rho}~,
\label{eq:scal}
\ee
as recognized first by Stephens~\cite{Stephens:1986ie}.  In fact, the
scaling solution satisfied the boundary conditions~\eqref{eq:bc1} and
\eqref{eq:bc2} and is expected to be similar to the full answer with
$\mu^2>0$.  In particular the scaling solution suggests that the
condensate may blow up as the acceleration increases.

With the conditions~\eqref{eq:bc1} and \eqref{eq:bc2} we can solve the
equation of motion~\eqref{eq:part} numerically and we show our
numerical results in Fig.~\ref{fig:num} for $\lambda=0.1$ and
$\lambda=0.01$.  We find that the condensate grows as $\rho$
decreases or as the proper acceleration $\alpha = 1/\rho$ increases.
Within the numerical accuracy as $\rho\to 0$, the condensate exhibits
diverging behavior similarly to the scaling solution.  This increasing
behavior of the condensate with acceleration reminds us of the
enhancement of the chiral condensate induced by the external magnetic
field, which is sometimes referred to as the magnetic catalysis.  So,
it should be appropriate to name the acceleration-induced enhancement
of the condensate \textit{acceleration catalysis}.

\section{Discussion and Conclusion}
\label{sec:concl}

Motivated by some discrepancies between the preceding results on the
condensate in an accelerated vacuum, we have revisited a real scalar
field theory in Rindler spacetime with a spontaneously broken $Z_2$
symmetry.  A key quantity for discussing a possible impact of the
acceleration on condensates is the Wightman two-point function in the
coincidence limit that represents quantum fluctuations.

First, we have studied the effect of acceleration on the two-point
Wightman functions for a free scalar field theory and clarified the
meaning of the choices of the vacuum and the observer.  This gives a
natural explanation for the observation that temperature-like
corrections in Rindler spacetime have a sign opposite to the genuine
thermal effect.  We have argued that such relations of the Wightman
function follow from the correspondence to thermo-field dynamics.

The most important part of this paper is the behavior of the scalar
condensate as a function of acceleration.  Based on our analysis we
can conclude that the condensate will not change as long as the system
is not accelerated regardless of where the observer sits, that is, the
condensate takes the vacuum value in agreement with
Ref.~\cite{Hill:1985wi,Hill:1986ec,Unruh:1983ac}.  This is non-trivial
in view of the fact that the Rindler observer perceives thermal
effects in the Minkowski vacuum regarding the particle distribution
that involves the energy dispersion relation.

Our analysis is, in principle, applied to such a system like a
superconductor placed on a transport craft with constant acceleration.
If a co-accelerated observer measures a condensate in this
superconductor, this observer should see that the condensate changes
depending on the acceleration.  What we found implies that the scalar
condensate increases with increasing acceleration.  We have named this
phenomenon acceleration catalysis.

We would stress that our main result differs from what is speculated
in some
papers~\cite{Ohsaku:2004rv,Kharzeev:2005iz,Ebert:2006bh,Lenz:2010vn,%
Castorina:2012yg,Takeuchi:2015nga}.  Ultimately, the fate of the
condensate depends on the definition of the coincidence limit of the
quantum fluctuation $\langle \hat{\phi}^2\rangle$ and the
regularization schemes.  Our assumption is that a finite deviation in
$\langle \hat{\phi}^2\rangle$ associated with acceleration should be
obtained by subtracting the common divergent pieces.  We would also
point out that we can in principle judge which of increasing and
descreasing scenarios should be the case using a Monte-Carlo
simulation on the lattice with non-trivial metric, and a preliminary
result favors our scenario of increasing condensate with larger
acceleration~\cite{yamamoto_priv}.

Although the above-mentioned subtraction procedure for acceleration
physics seems to be generally accepted, there are notable exceptions.
Dowker~\cite{Dowker:1978} advocated that the Minkowski vacuum
fluctuations should be regularized such that the Rindler vacuum
fluctuations are subtracted.  In our prescription this would
correspond to a co-accelerated observer making measurements while
taking the Rindler vacuum as a reference point.  This situation could
be a natural setting for the case of a black hole.  It is well-known
that Rindler spacetime is an approximation to Schwarzschild spacetime
in the near-horizon region.  Then, an observer at a fixed distance
from the horizon would find a positive thermal-like effect, provided
that the observer measures the condensate in the Minkowski vacuum
(corresponding to a freely falling frame).  Under such conditions it
seems conceivable that a condensate would melt as we approach the
black-hole horizon~\cite{Hawking:1980ng,Flachi:2011sx}.
When it comes to acceleration as opposed to well-established thermal
physics, the richness of physical outcomes from acceleration deserves
further attention.

\acknowledgments
This work was supported by MEXT-KAKENHI Grant No.\
24740169 (K.~F.) and by NEWFELPRO Grant Number 48 (S.~B.).
We acknowledge constructive remarks 
and discussions with Satoshi Iso.

\appendix
\section{Irrelevance of the observer}
\label{sec:app}

Here we articulate a step-by-step demonstration of independence of the
field expectation value regardless of the choice of the observer.

Let us first consider the situation with the Minkowski vacuum
$|M\rangle$.  Using the expanded form~\eqref{eq:phir2} we find,
\begin{align}
 &\langle M|\hat{\phi}(x)\hat{\phi}(x')|M\rangle \notag\\
 &= \int_0^\infty d\omega
 \int_0^\infty d\omega' \int d^2 k_\perp \int d^2 k_\perp'
 \Big( f_R f_R'\langle M|\hat{a}_R \hat{a}_R'|M\rangle \notag\\
 &\qquad + f_R f_R^{\prime\ast}\langle M|\hat{a}_R \hat{a}_R'^\dag|M\rangle
  + f_R^\ast f_R'\langle M|\hat{a}_R^\dag \hat{a}_R|M\rangle \notag\\
 &\qquad\qquad + f_R^\ast f_R^{\prime\ast}\langle M|
  \hat{a}_R^\dag \hat{a}_R^{\prime\dag}|M\rangle \Big) \;,
\label{eq:rfour}
\end{align}
where we used the expression for $\hat{\phi}$ quantized in Rindler
spacetime and we introduced a compact notation with the prime for
quantities with $\omega'$ and $\bk_\perp'$.  We plug
Eq.~\eqref{eq:therm1} and similar expressions for other combinations
of the creation/annihilation operators into Eq.~\eqref{eq:rfour}.  We
further replace $\omega$ with $-\omega$ to obtain:
\be
\begin{split}
 &\langle M|\hat{\phi}(x)\hat{\phi}(x')|M\rangle \\
 &= \frac{1}{2(2\pi)^4}
  \int\frac{dk^+}{2\pi k^+}\int d^2 k_\perp
  e^{i\bk_\perp\cdot(\bx_\perp-\bx_\perp')} \\
 & \times \Biggl[ \int_0^\infty \!\!d\omega\,
  F(\omega,k^+,\bk_\perp) \int_0^{\infty} \!\!d\omega'\,
  G(\omega',k^+,\bk_\perp)\\
 &\quad +\int_{-\infty}^0 \!\!d\omega\, F(\omega,k^+,\bk_\perp)
  \int_{-\infty}^0 \!\!d\omega'\, G(\omega',k^+,\bk_\perp)\Biggr]\;,
\end{split}
\label{eq:rfour2}
\ee
with the following functions that we define by
\begin{align}
 &F(\omega,k^+,\bk_\perp) \notag\\
 &\qquad\quad = e^{-i\omega\eta}\,
  e^{i\omega\log(k^+\sqrt{2}/\kappa)}
  K\Bigl(\omega,\frac{\kappa\rho}{2},\frac{\kappa\rho}{2}\Bigr)~, \\
 &G(\omega',k^+,\bk_\perp) \notag\\
 &\qquad\quad = e^{i\omega'\eta'}\,
  e^{-i\omega'\log(k^+\sqrt{2}/\kappa)}
  K\Bigl(\omega',\frac{\kappa\rho'}{2},\frac{\kappa\rho'}{2}\Bigr)~.
\end{align}
These are, of course, functions of $t$ and $z$ through $\eta$ and
$\rho$, which is not indicated in the argument of $F$ and $G$ for
notational brevity.  Also, in deriving Eq.~\eqref{eq:rfour2}, we
employed the analyticity property of the function
$K(\omega,\alpha,\beta)$.  Now we can deform the quantity in the angle
parentheses of Eq.~\eqref{eq:rfour2} as
\be
\begin{split}
 &\int_0^\infty d\omega F \int_0^\infty d\omega' G
 + \int_{-\infty}^0 d\omega F \int_{-\infty}^0 d\omega' G \\
 &\quad + \int_0^\infty d\omega F \int_{-\infty}^0 d\omega' G
 + \int_{-\infty}^0 d\omega F \int_0^\infty d\omega' G \\
 &= \int_{-\infty}^\infty d\omega F \int_{-\infty}^\infty d\omega' G~,
\end{split}
\ee
without changing its value;  two latter terms among four in the
left-hand side of the above expression give no contribution under the
$k^+$ integration.  After the $k^+$ integration, in fact, we can
easily show that the resulting contribution is proportional to
$\delta(\omega-\omega')$.  It is obvious that such a singularity at
$\omega=\omega'$ in Dirac's delta function cannot be picked up in the
integrals like $\int_0^\infty d\omega\int_{-\infty}^0 d\omega'$ and
$\int_{-\infty}^0 d\omega\int_0^\infty d\omega'$.

Eventually, we can rewrite the two-point function into the standard
form in Minkowski spacetime as
\begin{align}
 & \langle M|\hat{\phi}(x)\hat{\phi}(x')|M\rangle \notag\\
 &= \frac{1}{2(2\pi)^4}
  \int\frac{dk^+}{2\pi k^+} \int d^2 k_\perp\,
  e^{i\bk_\perp\cdot(\bx_\perp-\bx_\perp')} \notag\\
 &\qquad \times\int_{-\infty}^{\infty}d\omega\,
  F(\omega,k^+,\bk_\perp)\int_{-\infty}^{\infty}d\omega'\,
  G(\omega',k^+,\bk_\perp) \notag\\
 &=\int\frac{dk^+}{2k^+}\int d^2 k_\perp\,
  e^{i\bk_\perp\cdot(\bx_\perp-\bx_\perp')}
  e^{ik^+(x^--x^{-\prime})}e^{ik^-(x^+-x^{+\prime})}\;.
\label{eq:rfour3}
\end{align}
In this concrete process of calculations we note that we used
Eq.~\eqref{eq:usef} from the second line to the third line of
Eq.~\eqref{eq:rfour3}.

Next, we consider the situation with the Rindler vacuum $|R\rangle$.
We can find,
\be
\begin{split}
 &\langle R|\hat{\phi}(x)\hat{\phi}(x')|R\rangle \\
 &= \frac{1}{4\pi^4}
  \int_0^\infty d\omega \int d^2k_\perp\,
  e^{i\bk_\perp\cdot(\bx_\perp-\bx_\perp')} e^{-i\omega(\eta-\eta')} \\
 &\qquad\qquad\qquad\times
  \sinh(\pi\omega) K_{i\omega}(\kappa\rho)K_{i\omega}(\kappa\rho')
\end{split}
\label{eq:fourr1}
\ee
using the field operator in terms of the Rindler basis functions.  On
the other hand, we have,
\begin{align}
 &\langle R|\hat{\phi}(x)\hat{\phi}(x')|R\rangle = \int_0^\infty dk^+
 \int_0^\infty dk^{+\prime} \int d^2 k_\perp \int d^2 k_\perp' \notag\\
 &\quad\times \Bigl( f f'\langle R|\hat{a}_1 \hat{a}_1'|R\rangle
  + f f^{\prime\ast} \langle R|\hat{a}_1 \hat{a}_1'^\dag|R\rangle \notag\\
 &\quad\qquad + f^\ast f'\langle R|\hat{a}_1^\dag \hat{a}_1|R\rangle
  + f^\ast f^{\prime\ast}\langle R|
 \hat{a}_1^\dag \hat{a}_1'^\dag|R\rangle \Bigr)\;,
\label{eq:mfour}
\end{align}
where we introduced a compact notation; $f'$ and $\hat{a}_1'$ for
quantities with $x'$.  Using the definition~\eqref{eq:a21} and the
Bogolyubov transformation~\eqref{eq:bog} we have,
\be
\begin{split}
 & \langle R|\hat{a}_1(\bk_\perp,k^+)
  \hat{a}_1(\bk_\perp',k^{+\prime})|R\rangle \\
 & = -\delta^{(2)}(\bk_\perp+\bk_\perp') \frac{1}{k^+ k^{+\prime}} \\
 &\times \int_0^\infty\frac{d\omega}{2\pi}\,
  \alpha_\omega\beta_\omega\,e^{-i\omega\log(k^+\sqrt{2}/\kappa)}
  e^{i\omega\log(k^{+\prime}\sqrt{2}/\kappa')}~,
\end{split}
\label{eq:thermr1}
\ee
and similar expressions for the expectation values involving other
combinations of the annihilation ad creation operators.  Now we plug
these terms into Eq.~\eqref{eq:mfour} and use Eq.~\eqref{eq:bessi} to
define the $K$ function in a form of the $k^+$ and $k^{+\prime}$
integrations.  For example, we can show:
\be
\begin{split}
 &\int_0^\infty \frac{dk^+}{k^+}\, e^{ik^+ x^- + ik^- x^+}
  e^{-i\omega\log(k^+\sqrt{2}/\kappa)} \\
 &\qquad\qquad\qquad\qquad = K\Bigl(\omega,\frac{\kappa\rho}{2}e^\eta,
    \frac{\kappa\rho}{2}e^{-\eta}\Bigr)~.
\end{split}
\ee
Therefore, only the integrations over $\bk_\perp$ and $\omega$ remain.
Collecting all four terms, we finally find,
\begin{align}
 &\langle R|\hat{\phi}(x)\hat{\phi}(x')|R\rangle \notag\\
 &= \frac{1}{2(2\pi)^4}
  \int_0^\infty d\omega \int d^2 k_\perp\,
  e^{i\bk_\perp\cdot(\bx_\perp-\bx_\perp')} e^{-i\omega(\eta-\eta')} \notag\\
 &\qquad \times (\alpha_\omega^2 - 2\alpha_\omega\beta_\omega
  e^{-\pi\omega} + \beta_\omega^2 e^{-2\pi\omega}) \notag\\
 &\qquad \times
  K\Bigl(\omega,\frac{\kappa\rho}{2},\frac{\kappa\rho}{2}\Bigr)
  K\Bigl(\omega,\frac{\kappa\rho'}{2},\frac{\kappa\rho'}{2}\Bigr)~.
\label{eq:mfour1}
\end{align}
Using
$\alpha_\omega^2-2\alpha_\omega\beta_\omega
e^{-\pi\omega}+\beta_\omega^2e^{-2\pi\omega} = 1-e^{-2\pi\omega}$
together with Eq.~\eqref{eq:bessrel} we easily see that
Eq.~\eqref{eq:mfour1} is reduced to Eq.~\eqref{eq:fourr1}.

Now, finally, we want to sketch how this two-point function result can
be generalized for $n$-point ones.  We must understand that
$\hat{\phi}$ quantized in the Rindler vacuum is simply a restriction
of $\hat{\phi}$ on the Rindler right-wedge.  We can explicitly see
this from expressions of the type like Eq.~\eqref{eq:usef}.

Starting with the Minkowski quantized $\hat{\phi}$, therefore, we
have,
\begin{align}
 & \hat{\phi} = \int_0^{\infty}d\omega\int\frac{d^2k_\perp}{[2(2\pi)^4]^{1/2}}
  \biggl[ \hat{a}_2(\bk_\perp,\omega)
  K\Bigl(\omega,\frac{\kappa\rho}{2},\frac{\kappa\rho}{2}\Bigr) \notag\\
 &\qquad + \hat{a}_2^\dag(-\bk_\perp,-\omega)
  K^\ast\Bigl(-\omega,\frac{\kappa\rho}{2},\frac{\kappa\rho}{2}\Bigr)
  \biggr] e^{i\bk_\perp\cdot \bx_\perp-i\omega\tau} \notag\\
 &\quad + \int_0^{\infty}d\omega\int\frac{d^2k_\perp}{[2(2\pi)^4]^{1/2}}
  \biggl[ \hat{a}_2^\dag(\bk_\perp,\omega)
  K^\ast\Bigl(\omega,\frac{\kappa\rho}{2},\frac{\kappa\rho}{2}\Bigr)
  \notag\\
 &\qquad + \hat{a}_2(-\bk_\perp,-\omega)
  K\Bigl(-\omega,\frac{\kappa\rho}{2},\frac{\kappa\rho}{2}\Bigr)
  \biggr]e^{-i\bk_\perp\cdot \bx_\perp+i\omega\tau}~.
\label{eq:phi1}
\end{align}
By using the Bogolyubov transformations \eqref{eq:bog}, the
restriction to the right-wedge ($\rho>0$) gives $\hat{\phi}_R$, while
the restriction to the left-wedge ($-\rho>0$) gives $\hat{\phi}_L$.
Overall, we can write:
\be
 \hat{\phi} = \theta(\rho)\hat{\phi}_R + \theta(-\rho)\hat{\phi}_L \;.
\ee
Since the annihilation and creation operators from the right-wedge and
left-wedge mutually commute, and since left-wedge operators do
not act on $|R\rangle$ by definition, it follows that
\be
\begin{split}
 &\langle R|\hat{\phi}(x_1)\hat{\phi}(x_2)\dots\hat{\phi}(x_n)|R\rangle \\
 &\qquad\qquad
 = \langle R|\hat{\phi}_R(x_1)\hat{\phi}_R(x_2)\dots\hat{\phi}_R(x_n)|R\rangle~.
\end{split}
\ee


\end{document}